\documentclass[useAMS,usenatbib,usegraphicx]{mn2e}

\title[CDM power spectrum on sub-galactic scales] {The power spectrum
  of SUSY-CDM on sub-galactic scales} 
\author[A.~M.~Green, S.~Hofmann and D.~J.~Schwarz] 
  {Anne M.~Green$^{1,2,3}$\thanks{a.m.green@sheffield.ac.uk}, 
   Stefan Hofmann$^1$\thanks{stehof@physto.se} and 
   Dominik J.~Schwarz$^4$\thanks{dominik.schwarz@cern.ch} \\
   $^1$Physics Department, Stockholm University, AlbaNova University Center, SE-106 91 Stockholm, Sweden \\
   $^2$Astronomy Centre, University of Sussex, Brighton, BN1 9QH, UK\\
   $^3$Department of Physics and Astronomy, University of Sheffield,
       Sheffield, S3 7RH, UK\\
   $^4$Department of Physics, CERN, Theory Division, 1211 Geneva 23, 
       Switzerland}

\date{pre-print CERN-TH/2003-225, 23 September 2003}

\pagerange{\pageref{firstpage}--\pageref{lastpage}} \pubyear{2003}

\def\LaTeX{L\kern-.36em\raise.3ex\hbox{a}\kern-.15em
    T\kern-.1667em\lower.7ex\hbox{E}\kern-.125emX}

\begin{document}
\label{firstpage}
\maketitle

\begin{abstract}
The formation of large scale structure is independent of the nature of
the cold dark matter (CDM), however the fate of very small scale
inhomogeneities depends on the micro-physics of the CDM particles. We
investigate the matter power spectrum for scales that enter the Hubble
radius well before matter-radiation equality, and follow its evolution
until the time when the first inhomogeneities become non-linear. Our
focus lies on weakly interacting massive particles (WIMPs), and as a
concrete example we analyze the case when the lightest supersymmetric
particle is a bino. We show that collisional damping and
free-streaming of WIMPs lead to a matter power spectrum with a sharp
cut-off at about $10^{-6} M_\odot$ and a maximum close to that
cut-off. We also calculate the transfer function for the growth of the
inhomogeneities in the linear regime. These three effects (collisional
damping, free-streaming and gravitational growth) are combined to
provide a WMAP normalized primordial CDM power spectrum, which could
serve as an input for high resolution CDM simulations. The smallest
inhomogeneities typically enter the non-linear regime at a redshift of 
about~$60$.
\end{abstract}


\section{Introduction}

Analysis of the anisotropies in the cosmic microwave background (CMB)
radiation (Spergel et al. 2003) finds that the relative matter
density $\Omega_{{\rm m}} = 0.29 \pm 0.07$ is significantly larger
than the relative baryon density $\Omega_{{\rm b}} = 0.047 \pm
0.006$. These results are consistent with the observed abundances of
light elements and primordial nucleosynthesis (see e.g. Tytler et
al. (2000)) and the power spectrum found from galaxy red-shift surveys
(Percival et al. 2001), and indicate that the Universe contains a
significant amount of non-baryonic cold dark matter (CDM).

Weakly interacting massive particles (WIMPs) are attractive CDM
candidates, since a stable relic from the electroweak scale
generically has an interesting present day density, $\Omega_{\rm cdm}
\sim {\cal O}(1)$ (Dimopoulous 1990). The argument goes as follows:
annihilation processes cease and the WIMP number density $n$ becomes
fixed when the annihilation rate $\Gamma_{\rm ann}$ drops below the
expansion rate $H$ (usually referred to as chemical decoupling or
freeze-out). The temperature of chemical decoupling is thus defined by
$\Gamma_{\rm ann} = \langle \sigma_{\rm ann} v\rangle n(T_{\rm cd})
\sim H(T_{\rm cd})$, where $\sigma_{\rm ann}$ denotes the annihilation
cross section and $v$ the relative velocity of the annihilating
particles. For a typical weakly interacting particle one finds $T_{\rm
cd} \sim m/25$, $m$ being the WIMP mass. The present day relative WIMP
density can be estimated as
\begin{equation}
\Omega_{\rm wimp} = {m n_{\rm cd} (a_{\rm cd}/a_0)^3\over (3 H^2_0/8\pi G)} 
\sim 
0.2 {(m/T_{\rm cd})/25 \over \langle \sigma_{\rm ann} v\rangle/ 1 {\rm\ pb}}, 
\end{equation}
where $a$ denotes the scale factor and we have set $h = 0.7$ and
assumed $90$ relativistic degrees of freedom at chemical decoupling.
A cross section of about $1$~pb $(\equiv 10^{-40} {\rm m}^2)$ is typical 
for the annihilation of supersymmetric WIMPs since $1$ pb 
$\sim \alpha^2/(100 {\rm\ GeV})^2$, where $\alpha \approx 1/100$ is the 
electroweak coupling and supersymmetric particles typically have mass of 
order the electroweak scale ($100$ GeV).

In supersymmetry every standard model particle has an supersymmetric
partner and in most models there is a conserved quantum number
(R-parity), which makes the lightest supersymmetric particle
stable. Supersymmetry models have a large number of free parameters,
however in most models the lightest supersymmetric particle is the
lightest neutralino (which is a mix of the supersymmetric partners of
the photon, the $Z$ and the Higgs bosons; see e.g. Jungman, Kamionkowski 
\& Griest 1996), furthermore in large
regions of parameter space the lightest neutralino is mainly the bino
(Roszkowski 1991). Accelerator searches place a lower limit on the
neutralino mass of $m > 37$ GeV (see e.g. Hagiwara et al. 2002), while
the WMAP measurement of the matter density leads to an upper limit of
$m < 500$ GeV (Ellis et al. 2003).

In CDM cosmologies structure forms hierarchically; galaxy haloes form
from the merger and accretion of subhaloes (which themselves formed
from smaller subhaloes). The internal structure of galaxy haloes is
determined by the dynamical processes which act on the accreted
components. Dynamical friction causes subhaloes with mass $M > 10^{9}
M_{\odot}$ to spiral toward the centre of a Milky Way mass parent halo
within a Hubble time, while the tidal field of the parent halo can
strip material from a subhalo.  Numerically simulated galaxy haloes
contain large numbers of subhaloes that have not been destroyed
(Klypin et al. 1999; Moore et al. 1999) and the subhalo mass
function varies roughly as $N(M) \propto M^{-1.8} $ down to the
resolution limit of the simulations, $M \sim 10^6 M_{\odot}$ (Stoehr
et. al. 2003). Furthermore the anomalous flux ratios of multiply
imaged QSOs (Mao \& Schneider 1998; Schechter 2003) may be due to
millilensing by subhaloes, providing tentative observational
support for the existence of substructure in galaxy haloes.

Substructure on scales far smaller than those resolved by
numerical simulations has potentially significant consequences for
WIMP direct (Drukier, Freese \& Spergel 1986) and indirect (see
e.g. L. Bergstr\"om 2000) detection. WIMP direct detection (using
terrestrial detectors) probes the dark matter distribution on sub-milli-pc
scales (Silk \& Stebbins 1993; Moore et al. 2001; Green 2002, 2003).
WIMP indirect detection via WIMP annihilation products (gamma-rays,
antiprotons and neutrinos) is most sensitive to the highest density
regions of the Milky Way (e.g.  Silk \& Stebbins 1993; Bergstr\"om et
al. 1999, 2001; Calcaneo-Roldan \& Moore 2000) and clumping will also
enhance the extra-galactic gamma-ray signal (Ullio et al. 2002; Taylor
\& Silk 2003).  

In this letter we present the matter power spectrum close to the end
of the linear regime, for scales that enter the horizon well before
matter-radiation equality.  Three effects are important: after
chemical decoupling the primordial CDM density perturbations are
collisionally damped, due to elastic interactions with other species,
and then after kinetic decoupling free-streaming leads to further
damping. CDM inhomogeneities on physical scales larger than the
damping scale grow logarithmically during the radiation dominated
epoch and then roughly proportional to the scale factor during the matter
dominated epoch.

The small scale CDM power spectra for the density, velocity and
potential perturbations are an essential input for attempts to
analytically estimate the size and mass of the very first
gravitationally bound objects. Ultimately, very high resolution CDM
simulations will be necessary to make detailed predictions for the
fate of the very first objects, however the very different
characteristic smallest scales that have been predicted for different
CDM candidates [$10^{-6} M_\odot$ for neutralinos (Hofmann, Schwarz \&
St\"ocker 2001; Schwarz, Hofmann \& St\"ocker 2002; Berezinsky,
Dokuchaev \& Eroshenko 2003) compared with $10^{-13} M_\odot$ for
axions (Kolb \& Tkachev 1996)] could have observable astrophysical
consequences [such as femtolensing and picolensing of gamma ray bursts
(Gould 1992)]. This would open up the exciting possibility of
discriminating between CDM candidates astronomically.

\section{Collisional damping and free streaming}

Primordial CDM inhomogeneities on the smallest scales are smeared out
by collisional damping and free streaming, if the CDM has once been in
thermal contact with the hot component of the Universe (as is the case
for WIMPs). These damping processes give rise to a sharp cut-off in
the primordial CDM power spectrum. In order to make quantitative
predictions, we assume the CDM particle to be a bino with mass
$m$. The bino abundance is fixed at chemical decoupling, which happens
at typically $T_{\rm cd} \sim m/25$, i.e.~above $1$ GeV. Below that
temperature, elastic scattering of binos and fermions (leptons and
quarks from the radiation component) can still occur and the bino
decouples kinetically at significantly lower temperatures (Schmid,
Schwarz \& Widerin 1999):
$T_{\rm kd} = 10$ MeV to $40$ MeV,
depending on the parameters of the supersymmetric model.

Between chemical and kinetic decoupling the CDM particles interact
with the cosmic heat bath, which consists of all the relativistic
particles in the Universe, and this leads to collisional damping of
the CDM density perturbations. Since CDM is non-relativistic at this
epoch, the dominant processes are bulk and shear viscosity. Hofmann,
Schwarz \& St\"{o}cker (2001) calculated the effect of these
processes on the primordial density perturbations and found
exponential damping 
with a characteristic comoving wavenumber 
\begin{eqnarray}
k_{\rm d} 
&\approx & 
1.8  \left(\frac{m}{T_{{\rm kd}}}\right)^{1/2}\frac{a_{\rm kd}}{a_0}H_{\rm kd} 
\,, \nonumber \\
&\approx &   
\frac{3.8\times 10^7}{{\rm Mpc}}
\left({m\over 100 {\rm\ GeV}}\right)^{1/2} 
\left({T_{\rm kd}\over 30 {\rm\ MeV}}\right)^{1/2} \, .
\end{eqnarray}
This corresponds to a length scale of $\sim 10^{-2}/H$ at kinetic decoupling.
The total CDM mass contained in a sphere with radius $\pi/k_{\rm d}$ is 
$M_{\rm d} \sim 10^{-10} M_\odot$.

For a bino, the kinetic decoupling temperature depends on $m$ and on
the various sfermion masses $m_{\tilde{f}}$. Here we make the
simplifying assumption that all sfermions have the same mass, which is
not correct in more realistic models, 
but is a reasonable assumption for a first
estimate. We consider two fiducial sets of parameters: 
model A (B) has
$m = 100 (150)$ GeV and $m_{\tilde{f}} = 230 (190)$ GeV. For these
parameters we find, neglecting co-annihilations with other
supersymmetric particles, that chemical decoupling happens at $T_{\rm
cd} = 4.0 (5.8)$ GeV and $\Omega_{\rm bino} = 0.31 (0.17)$, thus models A
and B have CDM densities at the high and low end respectively of the
range of values found by WMAP.  For these models kinetic decoupling
occurs at $T_{\rm kd} = 33 (21)$ MeV and the damping mass scale is
$M_{\rm d} = 9 \times 10^{-11} (6 \times 10^{-11}) M_\odot$.

After kinetic decoupling, the CDM particles enter the free streaming regime. 
We calculate the effect of free streaming
by solving the collisionless Boltzmann equation
in Fourier space on subhorizon scales, to first order in the perturbed
thermodynamics quantities and neglecting metric perturbations, starting 
immediately after kinetic decoupling. We include the spectrum of CDM 
density perturbations at this time, and also take into account the matter 
and radiation components of the universe. The net result is again exponential 
damping with a characteristic scale which is inversely proportional to the 
(time-dependent) free-streaming length scale ($k_{{\rm fs}} \equiv 
2 \sqrt{3} / l_{{\rm fs}}$), multiplied by a polynomial term which arises 
from the ratio of the CDM particles kinetic energy to the thermal averaged 
kinetic energy. The free-streaming scale becomes approximately constant 
soon after matter-radiation equality;  
\begin{eqnarray}  
k_{\rm fs}
&\approx&
\left(\frac{m}{T_{\rm kd}}\right)^{1/2} {a_{\rm eq}/a_{\rm kd} \over
\ln\left(4a_{\rm eq}/a_{\rm kd}\right)} \frac{a_{\rm eq}}{a_0} 
H_{\rm eq} \,,
\nonumber \\
&\approx&
{1.7\times 10^6 \over {\rm Mpc}}
              {\left(m/100 {\rm\ GeV}\right)^{1/2}
               \left(T_{\rm kd}/30 {\rm\ MeV}\right)^{1/2}
              \over 1 + \ln\left(T_{\rm kd}/30 {\rm\ MeV}\right)/19.2}
\; .
\end{eqnarray}
The damping scale from free streaming depends on $\omega_{\rm m}\equiv
\Omega_{{\rm m}} h^2$ only via the logarithm, we therefore set it
equal to WMAP's best fit value, $\omega_{\rm m} = 0.14$ (Spergel et
al. 2003). The corresponding length scale at matter-radiation
equality $\sim 10^{-8}/H$ and the total mass contained in a sphere
with radius $\pi/k_{\rm fs}$ is $M_{\rm fs} \sim 10^{-6} M_\odot$.
More precisely, for model A (B) we find $M_{\rm fs} = 9 \times 10^{-7}
(6 \times 10^{-7}) M_\odot$.  

We can now put together all the damping factors for the CDM mass 
density perturbation $\Delta \equiv \delta \rho/\rho$.
Well after equality (our approximations are valid for $a/a_{\rm eq} \geq 10$),
when the comoving free streaming length is frozen, density inhomogeneities 
with large wavenumbers are suppressed by a damping factor 
\begin{equation}
D(k) =  
\left[1-\frac{2}{3}\left({k\over k_{\rm fs}}\right)^2\right]
{\rm exp}\left[-\left({k\over k_{\rm fs}}\right)^2-
\left({k\over k_{\rm d}}\right)^2\right].
\end{equation}
The present approximation is valid for $k/k_{\rm
fs} < 1$, as terms of order $(k/k_{\rm fs})^4$ have been neglected 
in the polynomial. 

The mass density contrast as function of redshift and wavenumber can 
now be written
\begin{equation}
\label{del1}
\Delta(k,z)
= \Delta(k,z_{\rm i}) T_{\Delta}^{\; 1/2}(k,z) D(k) \,,
\end{equation}
where $\Delta(k,z_{\rm i})$ is the primordial density perturbation
and $T_{\Delta}(k, z)$ is the transfer function which
encodes the gravitational evolution of $\Delta(k,z)$.  

\section{Gravitational growth on subhorizon scales}

Following Weinberg (2002), we model the Universe as a two component 
(non-relativistic matter and radiation) fluid. Here we utilize 
the zero shear (or 
longitudinal or conformal Newtonian) gauge. It is useful to define for 
each fluid (the index `a' stands for radiation,
$p_{\rm r} = \epsilon_{\rm r}/3$, or non-relativistic 
matter, $p_{\rm cdm} = 0$) the energy
density perturbation 
\begin{equation}
\Delta_{\rm a} = \frac{\delta\epsilon_{\rm a}}{\epsilon_{\rm a}+p_{\rm a}},
\end{equation}
(note that $\Delta_{{\rm cdm}} \equiv \Delta$) and the velocity
perturbations $v_{\rm a}$ by
\begin{equation}
{\bmath v}_{\rm a} = {{\rm i} {\bmath k}\over k}v_{\rm a},
\end{equation}
where ${\bmath v}_{\rm a}$ is the peculiar velocity.

The Newtonian gravitational potential $\phi$ is defined via the metric,
which is in the zero shear gauge given by 
\begin{equation}
\label{me}
{\rm d}s^2
= -\left(1+2\phi\right){\rm d}t^2 + a^2
\left(1-2\phi\right) \delta_{\rm ij}{\rm d}x^{\rm i}{\rm d}x^{\rm j} \,,
\end{equation}
in the absence of anisotropic stress,
where cosmic time is denoted by $t$.  Another useful quantity is the
hypersurface independent quantity $\zeta (= - {\cal R}) \equiv
\Delta/3 - \phi$ (Bardeen 1980, 1989), which is conserved on
superhorizon scales. We use this definition to make contact with the WMAP 
normalization, which is given in terms of ${\cal R}$ (Verde et al. 2003). 

Several comments are in order here. First, we work in the zero shear
gauge, because all subhorizon quantities can be interpreted in terms
of Newtonian physics, which is not the case in the synchronous gauge,
where the Newtonian gravitational potential is gauged to zero. 
Second, neutrinos are included in the radiation component in order to allow an
analytic treatment, i.e.\ their anisotropic stress is neglected.  This
leads to errors of around 10 \% (Hu et al. 1995). Furthermore we
ignore the effect of a non-zero cosmological constant or curvature
(which only effect the evolution of the perturbations at very late
times) and baryon inhomogeneities. At early times the baryons are 
tightly coupled to the radiation fluid, and photon diffusion damping 
rapidly erases small-scale perturbations in the baryon fluid at 
$z \sim 10^6$ to $10^5$. On small scales the tight coupling breaks down 
prior to recombination, and the baryon perturbations grow, however 
$\Delta_{\rm b} \ll \Delta_{\rm cdm}$ still (Yamamoto, Sugiyama and Sato 1997,
1998). Post decoupling on scales $k > k_{\rm b} \sim 10^{3} {\rm Mpc}^{-1}$
the residual electrons allow transfer of energy between the photon and 
baryon fluids so that thermal pressure prevents the baryon perturbations 
from growing, until $ z_{\rm b} \sim 150$ (Yamamoto, Sugiyama and Sato, 1997;
Padmanabhan, 2002). As we are interested in CDM perturbations on small scales 
at early times, we can neglect the perturbations in the baryon fluid.

The gravitational evolution of CDM inhomogeneities is described by the
corresponding transfer functions. For the calculation of the transfer
functions the equations of motion of the related perturbation
variables have to be solved. Unfortunately, this is not possible
exactly for all times, however there are two overlapping regimes for
which exact solutions exist. The first of these regimes is radiation
domination, $\epsilon_{{\rm r}} \gg \epsilon_{{\rm cdm}}$, for which an
exact solution can be found that is valid for all scales (Schmid,
Schwarz \& Widerin 1999). In the superhorizon limit ($k/a \ll H$),
$\phi \rightarrow \phi_0$, $\Delta_{\rm cdm, r} \to - 3\phi_0/2$,
$v_{\rm cdm, r} \to - (\phi_0/2) (k/aH)$ and $\zeta \to - 3\phi_0/2$,
while in the subhorizon limit ($k/a \gg H$) with $x = k/(\sqrt{3} a H)$:
\begin{eqnarray}
\label{sol1}
\Delta_{{\rm cdm}}(x) &=& - 9 \phi_{0} \left[
 \ln x + \gamma_{\rm E} -\frac{1}{2}\right] \,, \
v_{\rm cdm}(x) = - \frac{3 \sqrt{3} \phi_{0}}{x}   \,, \nonumber \\
\Delta_{\rm r}(x) &=& \frac{9 \phi_{0}}{2} \cos{x} \,, \
\phi(x) = - 3 \phi_{0} \frac{ \cos{x}}{x^2} \,,  
\end{eqnarray} 
with $\gamma_{\rm E}$ denoting Euler's constant.
We see that on sub-horizon scales during radiation
domination the matter perturbation $\Delta_{\rm cdm}$ grows
logarithmically while $\Delta_{\rm r}$ oscillates with constant
amplitude. Thus for small scales which enter the horizon sufficiently
long before matter-radiation equality the condition $\epsilon_{{\rm
cdm}} \Delta_{{\rm cdm}} \gg \epsilon_{{\rm r}} \Delta_{{\rm r}}$ is
satisfied during the radiation dominated era.
As long as the CDM density perturbations $\epsilon_{\rm cdm} \Delta_{\rm cdm}$ 
dominate the 
density perturbations of the radiation and baryons, the evolution of
$\Delta_{{\rm cdm}}$ is governed by (Hu and Sugiyama 1996):
\begin{equation}
\label{mes}
y(1+y) \frac{ {\rm d}^2 \Delta_{{\rm cdm}}}{ {\rm d} y^2}
         + \left(1 + \frac{3}{2} \right) \frac{ {\rm d} \Delta_{{\rm cdm}}}
            { {\rm d} y} - \frac{3}{2} (1- f_{\rm b}) \Delta_{{\rm cdm}} = 0 \,,
\end{equation}
where $y=a/a_{{\rm eq}}$ and $f_{\rm b}=\Omega_{\rm b}/ \Omega_{\rm m}$ is 
the baryon fraction, with best fit value from WMAP $f_{b}=0.16$ 
(Spergel et al. 2003). 
The exact solution to this equation is a combination 
of Legendre functions of first and second kind, $P_{\nu}(\sqrt{1+y})$ and 
$Q_{\nu}(\sqrt{1+y})$ with index $\nu(f_{b})=(\sqrt{25 - 24 f_{b}}-1)/2$. 
We smoothly join the small $y$ expansion of these functions with the 
radiation domination subhorizon solution [eq.~(\ref{sol1})], to obtain the 
normalisation of the solution to  
eq.~(\ref{mes}). Expanding the result for $y \gg 1$, we find
\begin{eqnarray}
\label{sol2}
\Delta_{\rm cdm}(y)= - 9 \phi_0\, c\,  y^{\nu/2}
\left[\ln\left(\frac{k}{k_{\rm eq}}\right) + b\right],
\end{eqnarray}
where $k_{\rm eq} = 1/(aH)_{\rm eq}$, 
$c(\nu) = \Gamma[1+2 \nu]/(2^\nu\Gamma^2[1+\nu])$ and
$b(f_{\rm b}) = 1/2 \ln(2^5/3) - \gamma_{\rm E} - 1/2 - 2/\nu - 
2\Gamma^\prime[\nu]/\Gamma[\nu]$, e.g.
$\nu(f_{\rm b}) = 1.80\, (2)$, $c[\nu(f_{\rm b})] = 1.37 (3/2)$ and 
$b(f_{\rm b}) = - 1.57 \, (-1.74)$ for $f_{\rm b} = 0.16 \, (0)$.
Note that before $z_{\rm b}$, CDM density perturbations
grow as $\Delta_{\rm cdm} \propto a^{\nu/2}$. Later the baryons follow the 
CDM and the matter fluctations grow as $a$. For the peculiar velocity and 
the Newtonian gravitational potential we obtain
\begin{eqnarray}
\label{sol2b}
v_{\rm cdm}(y) &=& \frac{k_{\rm eq}}{k} 
\sqrt{\frac{y}{2}} \frac{\rm d}{{\rm d} y}
\Delta_{\rm cdm}(y)
\; , \nonumber \\
\phi(y) &=& -\frac{3}{4} \left(\frac{k_{\rm eq}}{k}\right)^2 (1-f_{\rm b})
{\Delta_{\rm cdm}(y)\over y}.
\end{eqnarray}
In the following, we omit the subscript cdm.

For redshifts $z_{\rm eq} > z > z_{\rm b}$ (between matter-radiation 
equality and the epoch at which small-scale baryon perturbations start 
growing) we find the transfer function for the CDM density perturbations 
for modes which satisfy $k > k_{\rm b}$  
\begin{equation}
T_\Delta(k,z) = \textstyle{(6 c)^2 
\left[\ln\frac{k}{k_{\rm eq}} + b\right]^2
\left({1+z_{\rm eq}\over 1+z}\right)^\nu},
\end{equation}
and the transfer function for the Newtonian gravitational potential
on these scales is given by
\begin{equation}
T_\phi(k,z) = \textstyle{\left[\frac{27 (1-f_{\rm b}) c}{4}\right]^2       
\left[\ln\frac{k}{k_{\rm eq}} + b\right]^2
\left({k_{\rm eq}\over k}\right)^4  
\left({1+z_{\rm eq}\over 1+z}\right)^{\nu-2}\!\!\!}.
\end{equation}
The transfer function for the velocity depends on the initial time and
is therefore not a very useful quantity. 

\section{Power spectra}

In this section we present the dimensionless power spectra (defined as
${\cal P}_{X}(k,z) = (k^3/2\pi^2)\langle|X(k,z)|^2\rangle$) normalized
to the WMAP measurements (Spergel et al. 2003; Verde et al.
2003). For simplicity we assume that gravitational waves have a
negligible contribution to the CMB anisotropies and that the density
perturbations have a Harrison-Zel'dovich primordial power spectrum
($n=1$). We find for $k > k_{\rm b}$ and $z_{{\rm eq}} \gg z > z_{\rm b}$
\begin{eqnarray}
{{\cal P}_{\Delta}(k,z)\over 10^{-7} A} 
&=& \textstyle{
1.06\, c^2 \left[\ln\frac{k}{k_{\rm eq}} + b\right]^2 D^2(k)
\left(1+z_{\rm eq}\over 1+z\right)^\nu}, \\
{{\cal P}_v(k,z)\over 10^{-7} A} 
&=& 
0.13\, c^2 \nu^2
\nonumber \\
&& \hspace*{-1.5cm} 
\textstyle{\times
\left[\ln\frac{k}{k_{\rm eq}} + b\right]^2 
\left(\frac{k_{\rm eq}}{k}\right)^2 D^2(k)
\left(\frac{1+z_{\rm eq}}{1+z}\right)^{\nu-1},} \\
{{\cal P}_\phi (k,z)\over 10^{-7} A} 
&=&  
0.60\, c^2 (1-f_{\rm b})^2
\nonumber \\
&& \hspace*{-1.5cm}\textstyle{
\times\left[\ln\frac{k}{k_{\rm eq}} + b\right]^2 
\left(\frac{k_{\rm eq}}{k}\right)^4 D^2(k)
\left(\frac{1+z_{\rm eq}}{1+z}\right)^{\nu-2}},
\end{eqnarray}
where $A = 0.9 \pm 0.1$ according to Spergel et al. (2003). Note that
from the WMAP data $n = 0.99 \pm 0.04$, which is consistent with our
assumption of $n=1$.  The scale of equality is $k_{\rm eq} =
(0.01/{\rm Mpc}) (\omega_{\rm m}/0.14)$ and $1 + z_{\rm eq} = 3371
(\omega_{\rm m}/0.14)$.	

\begin{figure}
\setlength{\unitlength}{\linewidth}
\begin{picture}(1,0.66)
\put(0,-0.15){\includegraphics[width=\linewidth]{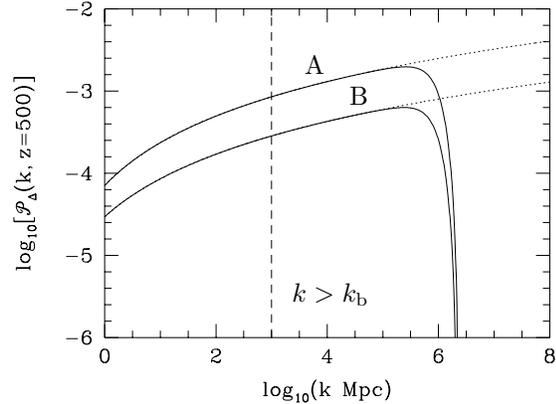}}
\put(0.45,0.15){\large $k > k_{\rm b}$}
\put(0.54,0.46){\large B}
\put(0.47,0.51){\large A}
\end{picture}
\caption{The dimensionless power spectrum of the CDM density contrast at 
$z = 500$ for models A and B from the text (full lines). Without 
the effects of collisional damping and free streaming, the power spectra
would be given by the dotted lines.\label{fig1}
}
\end{figure}

Fig.~\ref{fig1} shows the power spectrum for the density contrast at
a redshift of $500$, close to the end of the linear regime of
structure formation. It can be observed that the induced cut-off is
indeed very sharp and that the power spectrum has a maximum close to
the cut-off. 

\section{Discussion}

The mechanisms of collisional damping and free streaming of WIMPs lead
to a cut-off in the CDM power spectrum, which sets the typical scale
for the first haloes in the hierarchical picture of structure
formation. A rough estimate of the redshift at which typical fluctuations on
comoving scale $R$ go nonlinear can be made via
\begin{equation}
\label{znl}
\sigma(R, z_{\rm nl}) = 1, 
\end{equation}  
where $\sigma(R,z)$ is the mass variance defined by
\begin{equation}
\sigma^2(R, z) = \int_{0}^{\infty} W^2(kR) {\cal P}_{\Delta}(k,z)
        \frac{{\rm d} k}{k} \,,
\end{equation}
where $W(kR)$ is the fourier transform of the window function divided
by its volume. In accordance with the usual procedure, we take the
window function to be a top hat. For this calculation we need the
power spectrum on all scales, however our calculation of the CDM
transfer function is only valid for $k> k_{{\rm b}} \sim 10^{3} 
{\rm Mpc}^{-1}$. We therefore instead use the matter transfer function
found neglecting the baryon density (i.e. $f_{{\rm b}}=0$)
which is valid for $k > k_{{\rm eq
}}$ and take $\Omega_{{\rm m}} = 0.36 (0.22)$ for model A (B). This 
introduces errors at the $10 \%$ level, however the criteria for 
nonlinearity [eq.~(\ref{znl})] is only an order of magnitude estimate.
Finally we include the power spectrum on scales
$ k \sim k_{{\rm eq}}$ by normalizing $\sigma(R,z)$ to
$\sigma_8 \equiv \sigma(8/h\, {\rm Mpc},0) = 0.9 \pm 0.1$ (Spergel et
al. 2003), taking into account the suppression of the growth of
$\Delta$ at late times due to the cosmological constant.

\begin{figure}
\setlength{\unitlength}{\linewidth}
\begin{picture}(1,0.66)
\put(0,-0.15){\includegraphics[width=\linewidth]{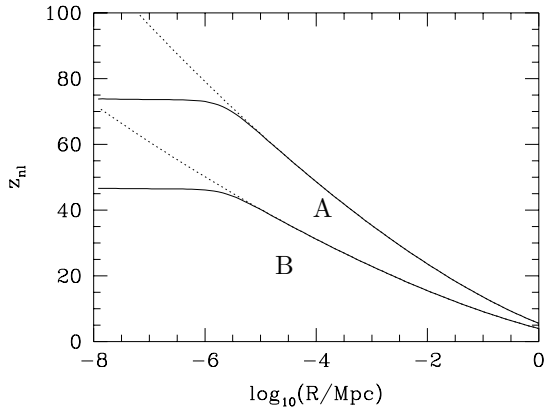}}
\put(0.5,0.29){\large A}
\put(0.44,0.2){\large B}
\end{picture}
\caption{The redshift at which typical fluctuations of comoving scale
$R$ become non-linear, for the two models discussed in the text. The
full lines take into account the effects of collisional damping and
free streaming, whereas the dashed lines show the
behaviour without a cut-off in the power spectrum. \label{fig2}}
\end{figure}

In fig. \ref{fig2} we plot $z_{\rm nl}$, as defined by
eq.~(\ref{znl}), as a function of the scale $R$.  The plateau at $R <
1$ pc is due to the sharp cut-off in the power spectrum.  We can now
give a more precise picture of the onset of the hierarchical structure
formation process; non-linear structure formation starts at a redshift
$z_{\rm nl}^{\max}$, which takes values in the range 30 to 80,
and $z_{\rm nl}^{\max} \sim 60$ for the best fit WMAP matter density.
To be more specific, this is the epoch when
the typical overdense regions go non-linear. Note, that rare
fluctuations with large amplitude will go non-linear at much higher
redshifts. If the density fluctuations have a gaussian probability
distribution then a $N \sigma$ fluctuation will go non-linear at
roughly $N z_{\rm nl}$. We leave the discussion of these rare large
fluctuations and their cosmological consequences for a forthcoming
paper and conclude with some comments on the typical fluctuations.

Let us estimate the size and mass of the first generation of subhaloes
that form at $z_{\rm nl}^{\max}$ using the spherical collapse model
(e.g. Padmanabhan 2002). The mean CDM mass within a sphere of comoving
radius R is $M(R)= 1.6 \times 10^{-7} M_{\odot}(\omega_{\rm m}/0.14)
(R/{\rm pc})^3$.  CDM overdensities that go non-linear have mass twice
this value i.e. roughly equal to the mass of Mars. These WIMP haloes
are however much less compact than Mars. The physical size of the
first haloes at turn-around (when their evolution decouples from the
cosmic expansion) is $r = 1.05 R/(1 + z_{\rm nl}^{\max}) \sim 0.02$
pc.  The first haloes then undergo violent relaxation, decreasing in
radius by a factor of two so that their present day radius is of order
tens of milli-pc (comparable to the size of the solar system). If some
of these first haloes could survive to the present day their
overdensity would be of order $\Delta_{{\rm halo}} = 7 (1+z_{\rm
nl}^{\max})^3 \sim 10^6$, several orders of magnitude larger than that
of galaxies.  Rare fluctuations, a non-gaussian fluctuation
distribution, or a blue ($n>1$) primordial power spectrum could lead
to even larger overdensities.

We regard this letter as a first step towards an ab-initio calculation
of the small scale structure of CDM. The resulting power spectra are
presented in a form that can be used as input for future very high
resolution simulations. Only once the fate of the first CDM haloes has
been understood in detail, will it be possible to make robust
predictions for the expected signals in direct and indirect dark
matter searches.

\section*{Acknowledgments}

We are grateful to Lars Bergstr\"{o}m, Joakim Edsj\"{o}, John Ellis, 
Eiichiro Komatsu, Ben Moore, Mia Schelke and Licia Verde 
for useful comments/discussions.
AMG was supported by the Swedish Research Council and the Particle Physics
and Astronomy Research Council (UK). SH was supported by the 
Wenner-Gren Foundation.

\label{lastpage}
\end{document}